\documentclass[12pt]{article}

\usepackage[margin=1in]{geometry}

\usepackage{graphicx}
\usepackage{float}
\usepackage{subfig}
\usepackage{hyperref}
\hypersetup{
    colorlinks=true,
    linkcolor=black,
    urlcolor=black,
    linktoc=all
}

\usepackage{amsmath} 
\usepackage{amssymb} 
\usepackage{bm} 

\newcommand{\comm}[2]{\left [ #1 ,  #2  \right]}
\newcommand{\sfrac}[2]{{\textstyle\frac{#1}{#2}}} 
\def\be{\begin{equation}}
\def\ee{\end{equation}}
\def\ba{\begin{eqnarray}}
\def\ea{\end{eqnarray}}
\def\bi{\begin{itemize}}
\def\ei{\end{itemize}}	
\def\l{\left}
\def\r{\right}
\def\fr{\frac}
\def\la{\label}
\def\pa{\partial}
\def\mpl{M_{\rm p}}
\def\lagr{\mathcal{L}}
\def\d{\mathrm{d}}

\usepackage{authblk} 
\usepackage[toc,page]{appendix} 

\numberwithin{equation}{section}

\newcommand{\eq}[1]{(\ref{#1})}

\begin{document} 
\title{\textbf{Anisotropic tensor modes \\from icosahedral inflation}}
\author{Jonghee Kang}
\author{Alberto Nicolis}
\affil{Center for Theoretical Physics and Physics Department,\\ \vspace{.15cm}
Columbia University, New York, NY 10027, USA}
\date{}

\maketitle 

\begin{abstract}
For icosahedral inflation, we compute the tensor modes' two-point function in the presence of higher derivative corrections, and show that in general this features anisotropies that are aligned with the underlying icosahedral structure. The effect is small within the regime of validity of the effective theory, but it is the leading contribution to a mixed correlator between the two different helicities. We also estimate the magnitude of a similar effect for a mixed scalar-tensor correlator, whose detailed computation we leave for future work.
Along the way, we clarify a number of aspects of the spin decomposition of generic icosahedral-invariant physical quantities.
\end{abstract}

\section{Introduction}

Icosahedral inflation \cite{Kang:2015aa} is a model for primordial inflation driven by an anisotropic solid with icosahedral symmetry. The icosahedral symmetry is powerful enough to force the background solution and the spectrum of scalar modes to be exactly isotropic, in agreement with observations, but it allows for  deviations from isotropy in scalar higher-point functions---starting with the three-point one---and in principle in the spectrum of tensor modes. We say `in principle' because in perturbation theory the leading origin of anisotropies for the tensor spectrum comes from higher-derivative corrections to the lowest order Lagrangian, which were not analyzed systematically in \cite{Kang:2015aa}. In a forthcoming paper \cite{us2} we identify which among the higher-derivative corrections have a chance of substantially impacting on the tensor modes' correlation functions while retaining a healthy effective field theory. The conclusion is that the non-minimal couplings between the matter fields and gravity should be perturbative w.r.t.~the lowest order Lagrangian at frequencies 
of order Hubble, with the exception of terms involving just one power of the Riemann tensor (possibly Ricci-contracted) coupled to the first derivatives of the matter fields---such terms can be arbitrarily large. As we will now argue, this implies that anisotropies in the tensor modes' two-point function have to be small\footnote{This is in contrast to what happens for e.g.~the scalar three-point function, where anisotropies can be large, so large as to overwhelm the isotropic signal.}.

Icosahedral inflation can be thought of as a theory of three scalar fields $\phi^I$ ($I=1,2,3$) coupled to gravity. The scalars enjoy standard shift symmetries,
\be
\phi^I \to \phi^I + a^I \; ,
\ee
as well as  internal icosahedral rotation symmetries,
\be
\phi^I = D^{I}_J \, \phi^J \; ,
\ee
where $D$ is any of the 60 rotation matrices making up the icosahedral group. To lowest order in derivatives, the action is
\be
S_0 = \int d^4 x \sqrt{-g} \big[ \sfrac12 \mpl^2 R + F(B^{IJ}) \big]  \; , \qquad B^{IJ} \equiv \pa_\mu \phi^I \pa^\mu \phi^J \; , 
\ee
where $F$ is invariant under the icosahedral group. The background values for the metric and scalars are
\be
ds^2 = -dt^2 + a^2(t) d \vec x \, ^2  \; , \qquad \phi^I = x^I \; ,
\ee
which break spatial translations and internal shifts down to their diagonal combinations, and spatial rotations and internal icosahedral rotations down to a diagonal icosahedral group. This means that the action for perturbations will be manifestly invariant under these diagonal icosahedral rotations. 

Let us focus on the tensor modes $\gamma_{ij}$. The building blocks for their quadratic, two-derivative action are 
\be
\dot \gamma_{ij} \dot \gamma_{kl} \; , \qquad \pa_i \gamma_{jk} \, \pa_l \gamma_{mn} \; ,\qquad \gamma_{ij} \gamma_{kl} \; ,
\ee
contracted with icosahedral invariant tensors. However, all icosahedral invariant two-index and four-index tensors are in fact fully isotropic \cite{Kang:2015aa}. At the six-index level, there is one anisotropic tensor $T^{ijklmn}_{\textrm{aniso}}$ that can be used to contract the indices of the structures above, but this means that anisotropies in the tensor modes' two-point function can only come from the gradient-energy term (which has six indices). But then, the lowest-order action above cannot yield such anisotropies, because the Einstein-Hilbert piece is fully isotropic, while $F$ does not involve derivatives of the metric. This is why we have to rely on higher derivative corrections.

These involve non-minimal couplings of our scalars to the Riemann tensor and its covariant derivatives, or higher covariant derivatives of our scalars, or both. From terms that involve one power of the Riemann tensor coupled to first derivatives of the scalars---the only terms that, as mentioned above, are not required to be perturbative at frequencies of order Hubble---we cannot get anisotropies in the tensor spectrum, simply because the Riemann only has four free indices. All other higher-derivative corrections have to be perturbative, and so anisotropies in the tensor modes' spectrum, if present, have to be small.

The goal of our paper is to compute the tensor modes' two-point function in the presence of these anisotropic corrections. Not surprisingly, we will find features in the direction-dependence of the two-point function that are aligned with the icosahedral geometry underlying our inflationary background. We will also find a nonzero mixed correlation function between the two different helicities, which is now compatible with the symmetries, since full $SO(3)$ invariance is broken in our model.

\section{Spin decomposition of icosahedral invariant tensors} \la{IcoTen}

In \cite{Kang:2015aa}, we showed that the space of icosahedral invariant six-index tensors is spanned by a basis made up of isotropic tensors (schematically of the form $\delta \delta \delta$) and of an anistropic one:
\be \la{Eq:IcoTen}
T^{ijklmn}_{\textrm{aniso}}  = 2(\gamma + 2) \, \delta^{ijklmn} + (\gamma + 1) \l( \delta^{ijkl}\delta^{mn}\delta^{m\, i+1} + \cdots \r) + \l( \delta^{ijkl}\delta^{mn}\delta^{m\, i-1} + \cdots \r) \; ,
\ee
where the dots stand for all other combinations of four and two indices out of six, the delta tensors with more than two indices are 1 if only if {\em all}  those indices take the same value, and $i+1$ and $i-1$ are to be interpreted modulo 3, that is $ 3+1 \sim 1$ and $ 1-1 \sim 3$. We further showed that the icosahedral invariant scalar 3-point function of icosahedral inflation can be decomposed into an isotropic part and a purely anisotropic one, in the sense that the overlap of the pure anisotropic part with any isotropic template vanishes. We used a Legendre polynomial expansion for that argument, but it turns out that one can directly decompose the tensor \eq{Eq:IcoTen} itself, and, not surprisingly, the symmetry properties of the 3-point function just follow from such a decomposition.

First, notice that \eq{Eq:IcoTen} is a totally symmetric tensor. A {\em generic} six-index spatial tensor can decomposed into irreps of $SO(3)$---from spin-0 to spin-6---as
\be
\mathbf{1} \otimes \mathbf{1} \otimes \mathbf{1} \otimes \mathbf{1} \otimes \mathbf{1} \otimes \mathbf{1} = 15 \cdot  \mathbf{0}\, \oplus \,  36 \cdot \mathbf{1} \, \oplus \, 40 \cdot \mathbf{2} \, \oplus \, 29 \cdot \mathbf{3} \, \oplus \, 15 \cdot \mathbf{4} \, \oplus \, 5 \cdot \mathbf{5} \, \oplus \, 1 \cdot \mathbf{6} \, .
\ee
However, upon totally symmetrizing, many such irreps are removed. Only one spin-0, one spin-2, one spin-4, and one spin-6 are left. Moreover, since the tensor \eq{Eq:IcoTen} is icosahedral invariant, so are its single traces and double traces. But all icosahedral invariant four-index and two-index tensors are isotropic \cite{Kang:2015aa}. In particular, given total symmetry, we must have
\be
T^{ijklmm}_{\textrm{aniso}} = A \l( \delta^{ij} \delta^{kl} + \delta^{ik}\delta^{jl} + \delta^{il}\delta^{jk} \r) \, , \qquad
T^{ijkkmm}_{\textrm{aniso}} = 5 A \, \delta^{ij} \; ,
\ee 
for suitable $A$. This implies that the spin-2 and spin-4 components of \eq{Eq:IcoTen} vanish, and we only have spin-0 and spin-6,
\be
T_\text{aniso}^{ijklmn} = \mathbf{0} \, \oplus \mathbf{6} \, .
\ee
The spin-0 component is
\be
T^{ijklmn}_{0} = \fr{ T_{\textrm{aniso}}^{iikkmm} }{105} S^{ijklmn} \, ,
\ee
where $S^{ijklmn} \equiv \l( \delta^{ij} \delta^{kl} \delta^{mn} + \text{14 other permutations}\r) $\footnote{The 105 is the analogue of the 3 we have when we decompose a two-index tensor into spin-0, spin-1, and spin-2: $T^{ij} = \fr{1}{3} T^{kk} \, \delta^{ij} + T^{[ij]} + \big( T^{(ij)} - \fr{1}{3} T^{kk} \,  \delta^{ij} \big)$.}, and so the tensor \eq{Eq:IcoTen} can be written as
\be \la{Eq:IcoTenDecom}
T^{ijkkmm}_{\textrm{aniso}} \equiv \fr{\l(\gamma+2\r)}{7}S^{ijklmn} + T^{ijklmn}_{6} \, , 
\ee
where $T^{ijklmn}_{6} $ represents the spin-6 component, which is traceless for all pairs of indices. Using this decomposition, we can reproduce the results of \cite{Kang:2015aa}. In particular, if we plug \eq{Eq:IcoTenDecom} into the last term of eq.~(4.20) of \cite{Kang:2015aa}, which is purely anisotropic when we set $\beta=8$, the terms involving our $S^{ijklmn}$  cancel the first three terms of that eq.~(4.20) exactly, leaving only a spin-6 trilinear interaction. 

Since we used only the transformation properties of \eq{Eq:IcoTen} under the rotation group, the argument we presented here is very general, so not only the scalar 3-point function, but any quantity involving \eq{Eq:IcoTen} can be decomposed into spin-0 and spin-6 components.
In fact, it can be shown that the icosahedral group is an isotropy subgroup of $SO(3)$ only for $SO(3)$-irreps of spin 0, 6, 10, 12, 15, 16, 18, and $\ge 20$, apart from 23 and 29
\cite{IHRIG19841}\footnote{A subgroup $H$ of $G$ is an isotropy subgroup for a given irrep $V$ of $G$, if a generic point in $V$ is invariant under $H$---that is, if the little group of a generic point in $V$ contains $H$. We are thankful to Austin Joyce for bringing ref.~\cite{IHRIG19841} to our attention.}. 
Rephrasing this result in our terms, this means that any quantity (function, tensor, etc.) that is invariant under the icosahedral group, will contain components only of spin 0, 6, 10, etc. Since the scalar three-point function of \cite{Kang:2015aa} is dominated by terms in the Lagrangian involving our six-index tensor above, it can only include {\em up to} spin 6, and so it includes spin 0 and 6. Were we to compute something that involved an icosahedral-invariant sixteen-index tensor, such as a scalar eight-point function, we would expect to find components of spin 0, 6, 10, 12, 15, and 16. And so on.

\section{The tensor modes' two-point functions} \label{PowerSpec}

As mentioned in the Introduction, we will treat the anistropic gradient energy of the tensor modes in a perturbative manner. Thus, the quadratic action is
\be \la{Eq:QuadLag1}
\begin{split} 
S_\gamma &= \fr{\mpl^2}{8} \int d^4 x \, a^3 Z \l[ \dot{\gamma}^2_{ij} - \fr{c^2_\gamma}{a^2} \l ( \pa_m \gamma_{ij}  \r )^2  -\fr{\Delta c_\gamma^2}{a^2}T^{ijklmn}_6 \pa_i \gamma_{jk} \pa_l \gamma_{mn} \r] \, ,
\\
& \equiv S_0 + S_\text{int} \, ,
\end{split}
\ee
where the free part is
\be
S_0 = \fr{\mpl^2}{8} \int d^4 x \, a^3 Z \l[ \dot{\gamma}^2_{ij} - \fr{c^2_\gamma}{a^2} \l ( \pa_m \gamma_{ij}  \r )^2 \r] \, ,
\ee
the interaction part is
\be
S_\text{int} = -\fr{\mpl^2}{8} \int d^4 x \, a^3 \, Z  \fr{\Delta c_\gamma^2}{a^2}T^{ijklmn}_6 \pa_i \gamma_{jk} \pa_l \gamma_{mn}  \, ,
\ee
and $\Delta c_\gamma^2$ is assumed to be small, and, to zeroth order in the slow-roll expansion, constant in time. $Z$ and $c^2_\gamma$ are also approximately constant in time, and in principle they can deviate substantially from one. We explore this possibility in a forthcoming paper \cite{us2}, but here, for notational simplicity, we will stick with 
\be
Z = c_\gamma^2 =1 \; .
\ee
Notice that, in writing \eqref{Eq:QuadLag1}, we have implicitly removed the spin-0 component of our tensor \eqref{Eq:IcoTenDecom} by reabsorbing it into the isotropic gradient energy. This just amounts to redefining $c_\gamma^2$. We have also neglected non-derivative terms, since they are suppressed by slow-roll parameters \cite{Endlich:2012pz}.

In perturbation theory, the corrections to the two-point functions can be computed in a manner similar to the way in which we compute  higher-point correlation functions.
To this end, we rewrite our Lagrangian in spatial Fourier space and conformal time:
\begin{align} 
S_\gamma = \fr{\mpl^2}{8} \int \fr{d \tau d^3 k}{(2\pi)^3} & \,  a^2 \Big[ \gamma'_{ij}(\vec{k}, \tau) \gamma'_{ij}(-\vec{k}, \tau) -k^2 \gamma_{ij}(\vec{k}, \tau) \gamma_{ij}(-\vec{k}, \tau) \nonumber \\
& - \Delta c_\gamma^2 \, k_i k_l \, T^{ijklmn}_6 \gamma_{jk}(\vec{k}, \tau) \gamma_{mn}(-\vec{k}, \tau) \Big] \, . \la{Eq:QuadLag2}
\end{align}
The conventions we use are the same as Maldacena's \cite{Maldacena:2002vr}:
\begin{gather}
\gamma_{ij}(\vec k,\tau) = \sum_{s=\pm} \gamma^s (\vec{k},\tau) \, \epsilon_{ij}^s(\vec k)\\
\epsilon^s_{ii} = k_i \epsilon^s_{ij} = 0 \, , \quad \epsilon^s_{ij} \epsilon^{s'*}_{ij} = 2 \delta^{ss'}  \label{epsilon}
\\
\gamma^s (\vec{k},\tau) = \gamma_{cl}(k,\tau) a^s (\vec{k}) + \gamma_{cl}(k,\tau)^* a^{s\dagger}(-\vec{k}) 
\\
\big[ a^s(\vec{k}), \, a^{s'\dagger}(\vec{k}')\big] = (2\pi)^3 \delta^3(\vec{k} - \vec{k}') \delta^{ss'} \, ,
\end{gather}
where the classical solution $\gamma_{cl}(k,\tau)$ obeys the equation of motion obtained by varying $S_0$,
\be
\fr{d^2}{d\tau^2} \gamma_{cl} + 2aH \fr{d}{d\tau} \gamma_{cl} + k^2 \gamma_{cl}=0 \, ,
\ee
which, to zeroth order in slow-roll, yields
\be \la{eq:gammacl}
\gamma_{cl} (k,\tau) = \fr{1}{\mpl a} \fr{e^{-ik\tau}}{\sqrt{k}} \l( 1 - \fr{i}{k\tau} \r) \, .
\ee


Following standard perturbation theory for cosmological correlation functions, the correction to the $++$ two-point function is
\be
\delta \langle \gamma^+(\tau)^2 \rangle = -i \int_{-\infty}^\tau d \tau' \, \l \langle \Omega(-\infty) | \comm{\gamma^+(\tau)^2}{H_\text{int} (\tau')} | \Omega(-\infty) \r \rangle \, ,
\ee
where $H_\text{int} = - \int d^3 x \lagr_\text{int}$. To be more explicit, defining $\gamma^s_i \equiv \gamma^s(\vec{k}_i,\tau)$,
\be \la{eq:2ptpp}
\begin{split}
\delta \l \langle \gamma^+_1 \gamma^+_2 \r \rangle &=  -\fr{i \mpl^2 \Delta c_\gamma^2}{8} \int_{-\infty}^\tau \fr{d \tau' d^3 k'_1 d^3 k'_2}{(2\pi)^3} \, \delta^3(\vec{k}_1' + \vec{k}_2') \, a^2   T^{ijklmn}_6  k'_{1i} k'_{1l} \epsilon^+_{jk}(\vec{k}_1') \epsilon^+_{mn}(-\vec{k}_1')  
\\
& \quad \times \l \langle \comm{\gamma^+_1 \gamma^+_2}{\gamma^+_{1'} \gamma^+_{2'}} \r \rangle \, ,
\end{split}
\ee
where  the `primed' $\gamma$'s are evaluated at $\tau'$.
%
After, some straightforward algebra, this becomes
\be
\delta \l \langle \gamma^+_1 \gamma^+_2 \r \rangle =   -\fr{ \mpl^2 \Delta c_\gamma^2}{4} (2\pi)^3 \delta^3(\vec{k}_1 + \vec{k}_2) \times  T^{ijklmn}_6 \,  k_{1i} k_{1l}  \, \epsilon^+_{jk}(\vec{k}_1) \epsilon^+_{mn}(-\vec{k}_1)   \times I(\tau;\, -\infty) \, ,
\ee
where
\be
I(\tau_1; \, \tau_2) \equiv \Big[ i \gamma_\text{cl}(k_1,\tau_1) \gamma_\text{cl}(k_2,\tau_1) \int_{\tau_2}^{\tau_1} \d \tau' \, a^2 \gamma_\text{cl}(k_1,\tau_2)^* \gamma_\text{cl}(k_2,\tau_2)^* \Big]+ \rm{c.c.} 
\ee
Substituting \eq{eq:gammacl} into the above integral, we obtain
\be
I(\tau ;\,-\infty) = \fr{2}{\mpl^4 a^2}\fr{1}{2k_1^5 \tau^2} \l(k_1^2 \tau^2 + 3\r) \, .
\ee
As a result, the correction to the two-point function  at late times is
\be \la{eq:2ptpp_done}
\begin{split}
\delta \l \langle \gamma^+_1 \gamma^+_2 \r \rangle &= -(2\pi)^3 \delta^3(\vec{k}_1 + \vec{k}_2) T^{ijklmn}_{6} \hat{k}_{1i} \hat{k}_{1l}  \epsilon^+_{jk}(\vec{k}_1) \epsilon^+_{mn}(-\vec{k}_1)  \fr{H^2 }{\mpl^2} \fr{3 \Delta c_\gamma^2}{4k_1^3} \, .
\end{split}
\ee
Thanks to parity, for $ \l \langle \gamma^-_1 \gamma^-_2 \r \rangle$ we get the same result:
\begin{align} \la{eq:2ptmm_done}
\delta \l \langle \gamma^-_1 \gamma^-_2 \r \rangle & = -(2\pi)^3 \delta^3(\vec{k}_1 + \vec{k}_2) T^{ijklmn}_{6} \hat{k}_{1i} \hat{k}_{1l}  \epsilon^-_{jk}(\vec{k}_1) \epsilon^-_{mn}(-\vec{k}_1)  \fr{H^2 }{\mpl^2} \fr{3 \Delta c_\gamma^2}{4k_1^3} \\
& = \delta \l \langle \gamma^+_1 \gamma^+_2 \r \rangle 
\end{align}

Finally, we can study mixed $+-$ correlation functions: these vanish exactly in the absence of anisotropies, and thus provide perhaps the most direct observational window on our model. 
Using the same method as above, we get
\be \la{eq:2ptmp_done}
\begin{split}
\delta \l \langle \gamma^-_1 \gamma^+_2 \r \rangle &= -(2\pi)^3 \delta^3(\vec{k}_1 + \vec{k}_2) \,  T^{ijklmn}_{6} \hat{k}_{1i} \hat{k}_{1l}  \epsilon^+_{jk}(\vec{k}_1) \epsilon^-_{mn}(-\vec{k}_1) \,  \fr{H^2 }{\mpl^2} \fr{3 \Delta c_\gamma^2}{4k_1^3} \, .
\end{split}
\ee
Similarly, $\delta \l \langle \gamma^+_1 \gamma^-_2 \r \rangle$ is
\begin{align} \la{eq:2ptpm_done}
\delta \l \langle \gamma^+_1 \gamma^-_2 \r \rangle &= -(2\pi)^3 \delta^3(\vec{k}_1 + \vec{k}_2) T^{ijklmn}_{6} \hat{k}_{1i} \hat{k}_{1l}  \epsilon^-_{jk}(\vec{k}_1) \epsilon^+_{mn}(-\vec{k}_1)  \fr{H^2 }{\mpl^2} \fr{3 \Delta c_\gamma^2}{4k_1^3} \\ 
& = \delta \l \langle \gamma^-_1 \gamma^+_2 \r \rangle^*
\end{align}

All these results can be summarized compactly as
\be \label{compact}
\langle \gamma_1^s \gamma_2^{s'}  \rangle = \big[ \delta^{ss'} - \sfrac{3}{4} \Delta c_\gamma^2   \, M^{ss'} (\vec k_1) \big] \, \langle \gamma_1 \gamma_2 \rangle_0  \; ,
\ee
where $M$ is a direction-dependent $2\times2$ matrix in polarization space,
\be \label{M}
M^{ss'}(\vec k) \equiv T^{ijklmn}_6 \, \hat{k}_{i} \hat{k}_{l} \,   \epsilon^{s}_{jk}(-\vec{k}) \epsilon^{s'}_{mn}(\vec{k})  \;, 
\ee
and $\langle \gamma_1 \gamma_2 \rangle_0$ is the standard two-point function for tensor modes,
\be
\langle \gamma_1 \gamma_2 \rangle_0 \equiv (2\pi)^3 \delta^3(\vec{k}_1 + \vec{k}_2) \fr{H^2 }{\mpl^2} \fr{1}{k_1^3}  \; .
\ee

\section{A closer look at anisotropies}

From eq.~\eqref{compact}, we see that the matrix $M$ encodes all the directional dependence of our two-point functions. 
Given that our $T_6$ tensor transforms in a spin-6 representation of rotations, one would hope that the entries of $M$, as functions of $\hat k$,  can be expanded in $\ell = 6$ spherical harmonics. This is  certainly true for $M^{++}$ and $M^{--}$. Using the identities of Appendix \ref{identities} and the tracelessness and total symmetry of $T_6$, we get
\begin{align}
M^{++}(\vec{k}) = M^{--} (\vec{k}) & = \sfrac12 \, T^{ijklmn}_6 \, \hat{k}_{i} \hat{k}_{j} \hat{k}_{k} \hat{k}_{l} \hat{k}_{m} \hat{k}_{n} \; ,  \\
& =  \sum_{m=-6}^6 A_{6m} Y_6^m (\theta, \phi) \ \label{M++}
\end{align}
with
\be
A_{6, \pm 6} = -  \sqrt{\sfrac{5}{11}} \cdot  A_{6, \pm 2} = - \sfrac52  \gamma \sqrt{\sfrac{3\pi}{1001}}  \; , \qquad A_{6 ,\pm 4} =  \sqrt{\sfrac{7}{2}} \cdot  A_{6, \pm 0} = ( \gamma+2) \sqrt{\sfrac{\pi}{182}} \; .
\ee

However, for $M^{+-}$ and $M^{-+}$ we have to be more careful. If we try to expand them in spherical harmonics, we get contributions from arbitrarily high $\ell$'s. What's going on? The subtlety has to do with an arbitrariness implicit in the definition of the polarization tensor $\epsilon^s_{ij}(\vec k)$: its tensor structure is uniquely determined by its helicity and by \eqref{epsilon}, but its phase is  a matter of definition; moreover, as a function of $\vec k$, such a phase has singularities, which can be moved around, but not removed completely. We explore and clarify this in  Appendix \ref{phases}. To reproduce those singularities in a spherical harmonics expansion, one needs all angular momenta up to $\ell = \infty$. This is not a problem for $M^{++}$ and $M^{--}$, because there the phase of $\epsilon^s_{ij}(\vec k)$ cancels against that of  $\epsilon^{s}_{ij}(-\vec k) = \epsilon^{s \, *}_{ij}(\vec k)$, leaving us with completely unambiguous and regular functions of $\vec k$. But for $M^{+-}$ and $M^{-+}$ the cancellation does not happen, and we are left with ambiguous and singular functions of $\vec k$.

To circumvent the problem, we eliminate the offending phase by considering the absolute value squared of our functions. This is unambiguous and regular. The small price to pay is that now we need spherical harmonics up to $\ell = 12$ instead of $6$
\footnote{We cannot hope to get back down to $\ell=6$ by considering the absolute value rather than the absolute value squared: the latter is a bilinear form, for which the addition of angular momenta is simple, but the former is the square root of a bilinear form,  which  would take us again all the way up to $\ell = \infty$.}. 
In fact, given the discussion at the end of sect.~\ref{IcoTen}, we expect contributions only from $\ell=0,6,10,12$. 
Indeed, using again the identities of Appendix \ref{identities} and the tracelessness and total symmetry of $T_6$, we find
\begin{align}
| M^{+-}(\vec{k}) |^2 = | M^{-+} (\vec{k})  |^2 & = \sfrac12 \big( | M^{+-}(\vec{k}) |^2 + | M^{-+}(\vec{k}) |^2 \big) \nonumber \\
& = 2 \, T^{ijklmn}_6 T^{opqrst}_6 \,  \hat{k}_{i} \hat{k}_{j} \hat{k}_{o} \hat{k}_{p} P^{TT}_{(klmn),(qrst)} \nonumber \\
&=    \sum_{\ell, m} B_{\ell m} Y_\ell^m (\theta, \phi) \, , \label{M+-}
\end{align}
(the eight-index $P^{TT}$ projector is defined in Appendix \ref{identities}), with
\begin{align}
\mbox{$\ell = 0$:} \quad  & B_{0,0}  = \sfrac{1280 \sqrt{\pi}}{1001} (\gamma +1) \\ 
\mbox{$\ell = 6$:} \quad & B_{6, \pm 6} = \sqrt{\sfrac{5}{11}} \cdot  B_{6, \pm 2} = \sfrac{160}{323}  \sqrt{\sfrac{3 \pi}{1001}} (3 \gamma+1) \nonumber \\ 
& B_{6, \pm 4} = \sqrt{\sfrac{7}{2}} \cdot B_{6, 0} = \sfrac{160}{323}  \sqrt{\sfrac{2 \pi}{91}} (\gamma+1) \\
\mbox{$\ell = 10$:} \quad  & B_{10, \pm 10} = \sqrt{\sfrac{255}{19}} \cdot  B_{10, \pm 6} = \sqrt{\sfrac{255}{494}} \cdot B_{10, \pm 2} = \sfrac{415}{23}  \sqrt{\sfrac{3 \pi}{323323}} (3 \gamma+1) \nonumber \\
&  B_{10, \pm 8} = \sfrac12 \sqrt{\sfrac{17}{3}} \cdot B_{10, \pm 4} = \sqrt{\sfrac{187}{130}} \cdot  B_{10, 0} = \sfrac{415}{23}  \sqrt{\sfrac{10 \pi}{51051}} (\gamma+1) \\\
\mbox{$\ell = 12$:} \quad  & B_{12, \pm 12} = 5 \sqrt{\sfrac{69}{154}} \cdot  B_{12, \pm 8} = \sfrac{15}{17}\sqrt{\sfrac{437}{187}} \cdot B_{12, \pm 4} = \sfrac{5}{58}\sqrt{\sfrac{5681}{119}} \cdot B_{12, 0}= \sfrac{45}{4}  \sqrt{\sfrac{\pi}{676039}} (\gamma+1) \nonumber \\
&  B_{12, \pm 10} = \sfrac15 \sqrt{\sfrac{209}{21}} \cdot B_{12, \pm 6} = \sqrt{\sfrac{209}{34}} \cdot  B_{12, \pm 2} = \sfrac{33}{46}  \sqrt{\sfrac{3 \pi}{29393}} (3\gamma+1)
\end{align}

\begin{figure}[H]
\centering
\subfloat{\includegraphics[width=5.5cm]{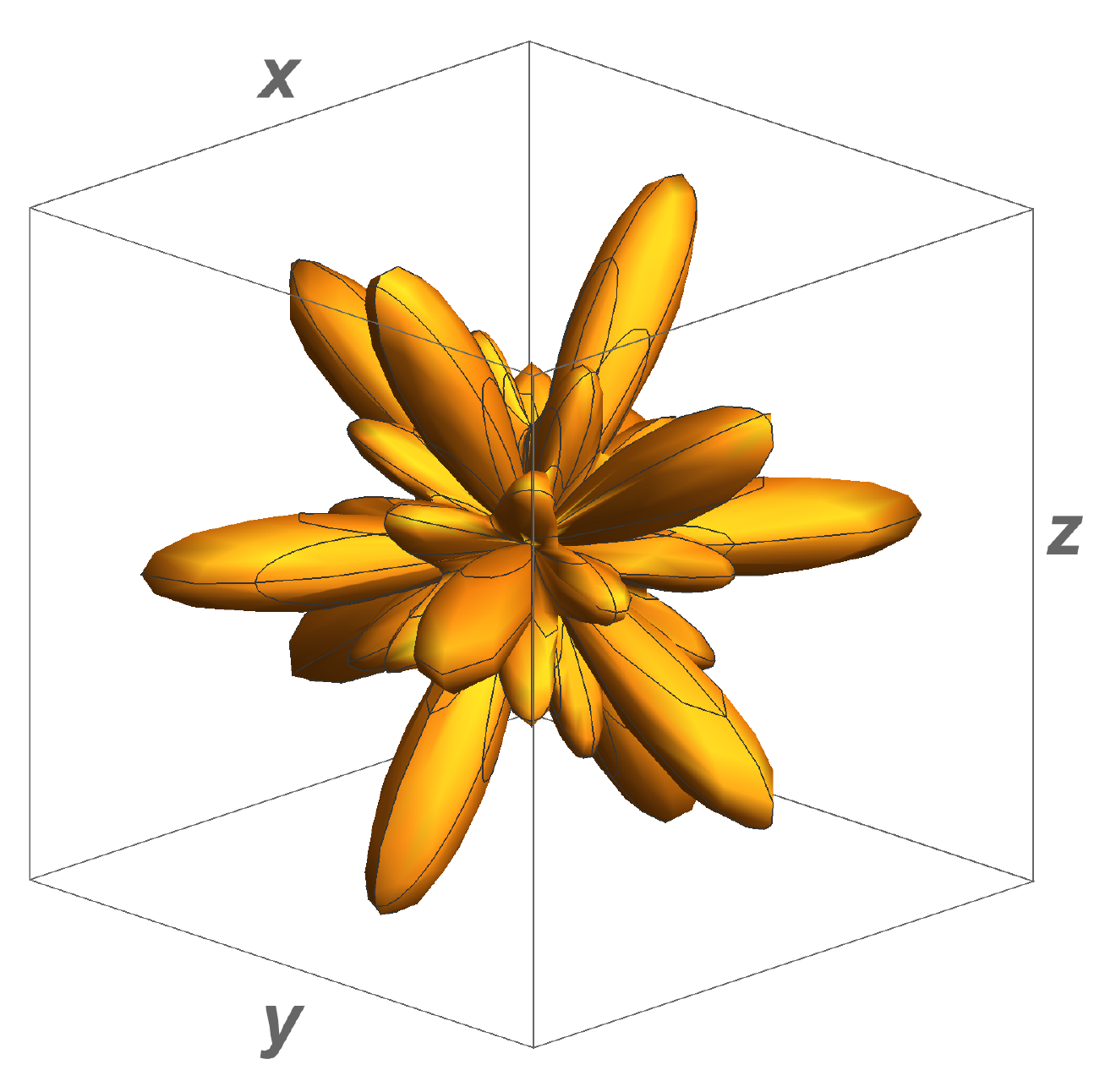}  
} \hspace{1cm}
\subfloat{\includegraphics[width=5.5cm]{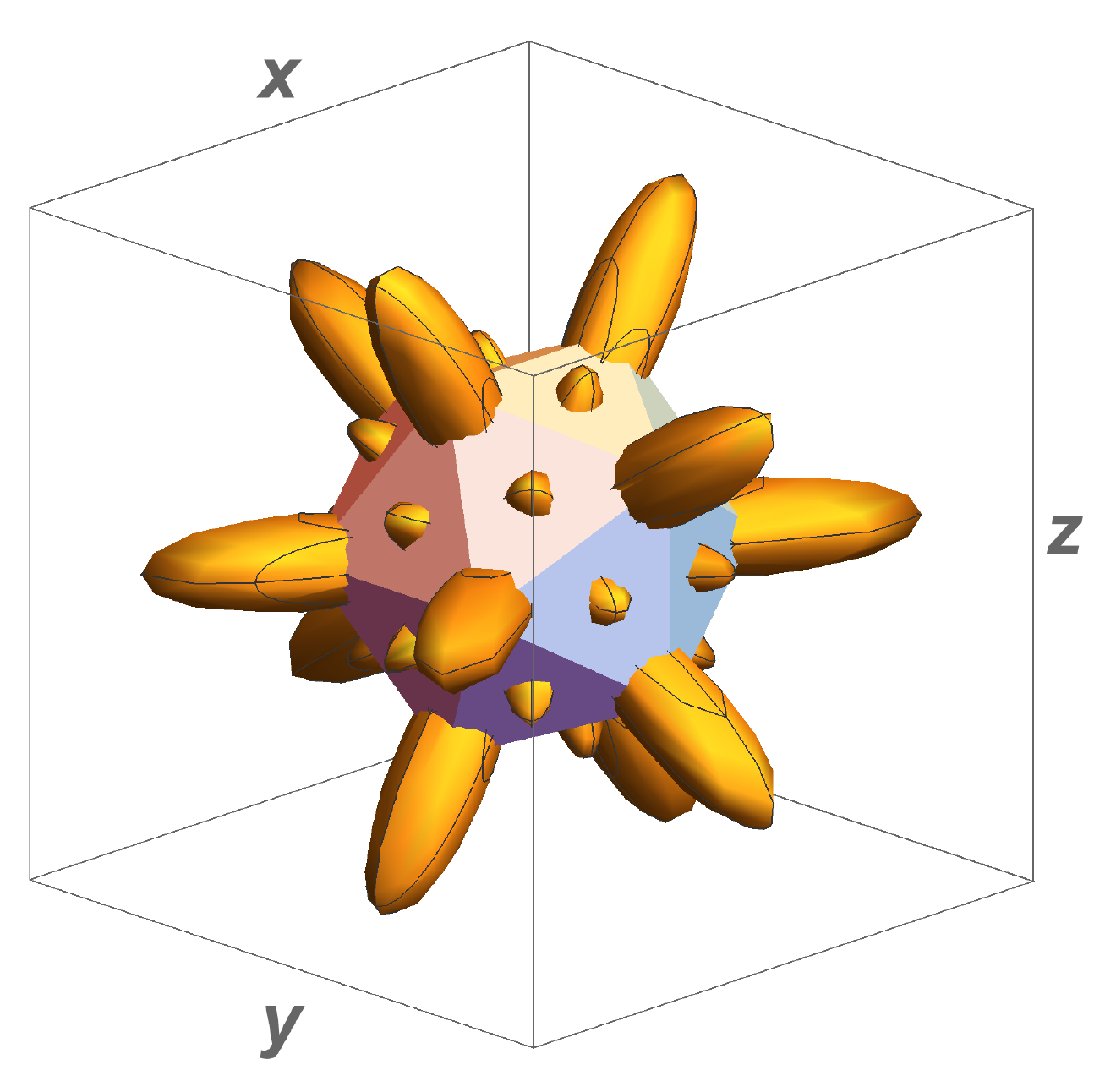} 
}
\caption{\label{++} \small $M^{++}(\hat{k}) = M^{--} (\hat{k})$.}
\end{figure}

\begin{figure}[H]
\centering
\subfloat{\includegraphics[width=5.5cm]{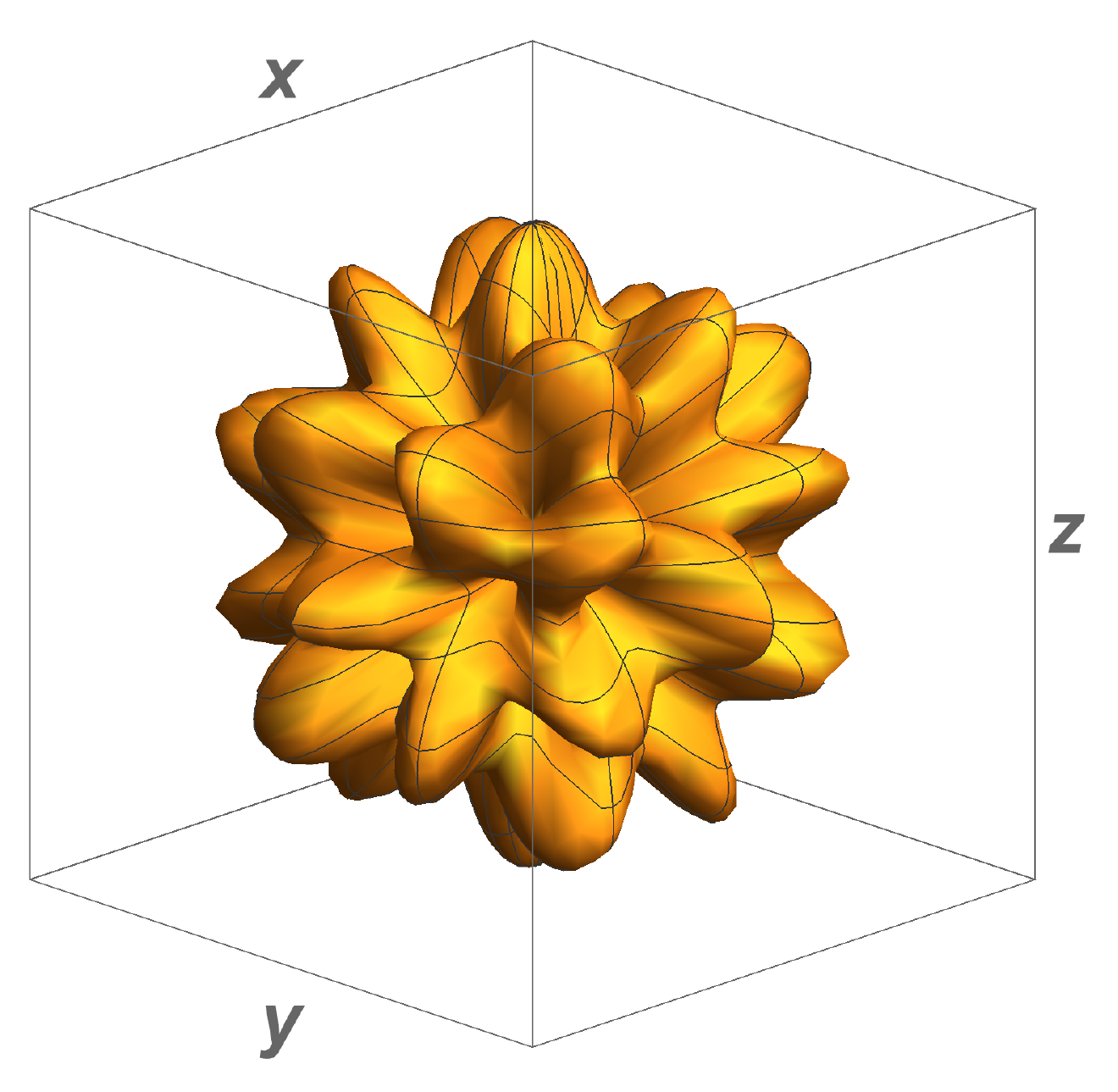}  
} \hspace{1cm}
\subfloat{\includegraphics[width=5.5cm]{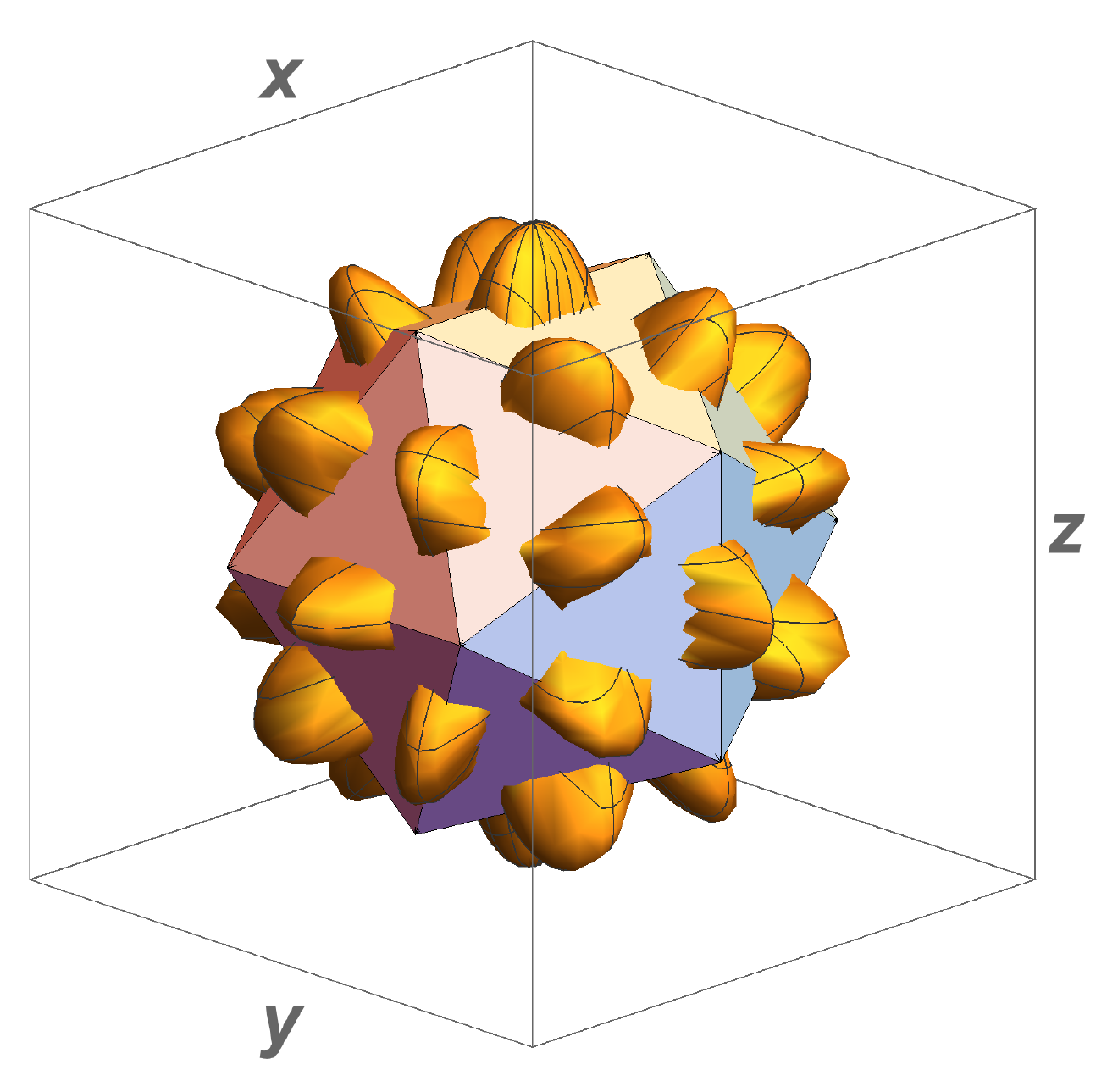} 
}
\caption{\label{+-} \small $|M^{+-}(\hat{k})| = |M^{-+} (\hat{k})|$.}
\end{figure}

\begin{figure}[H]
\centering
\subfloat{\includegraphics[width=5.5cm]{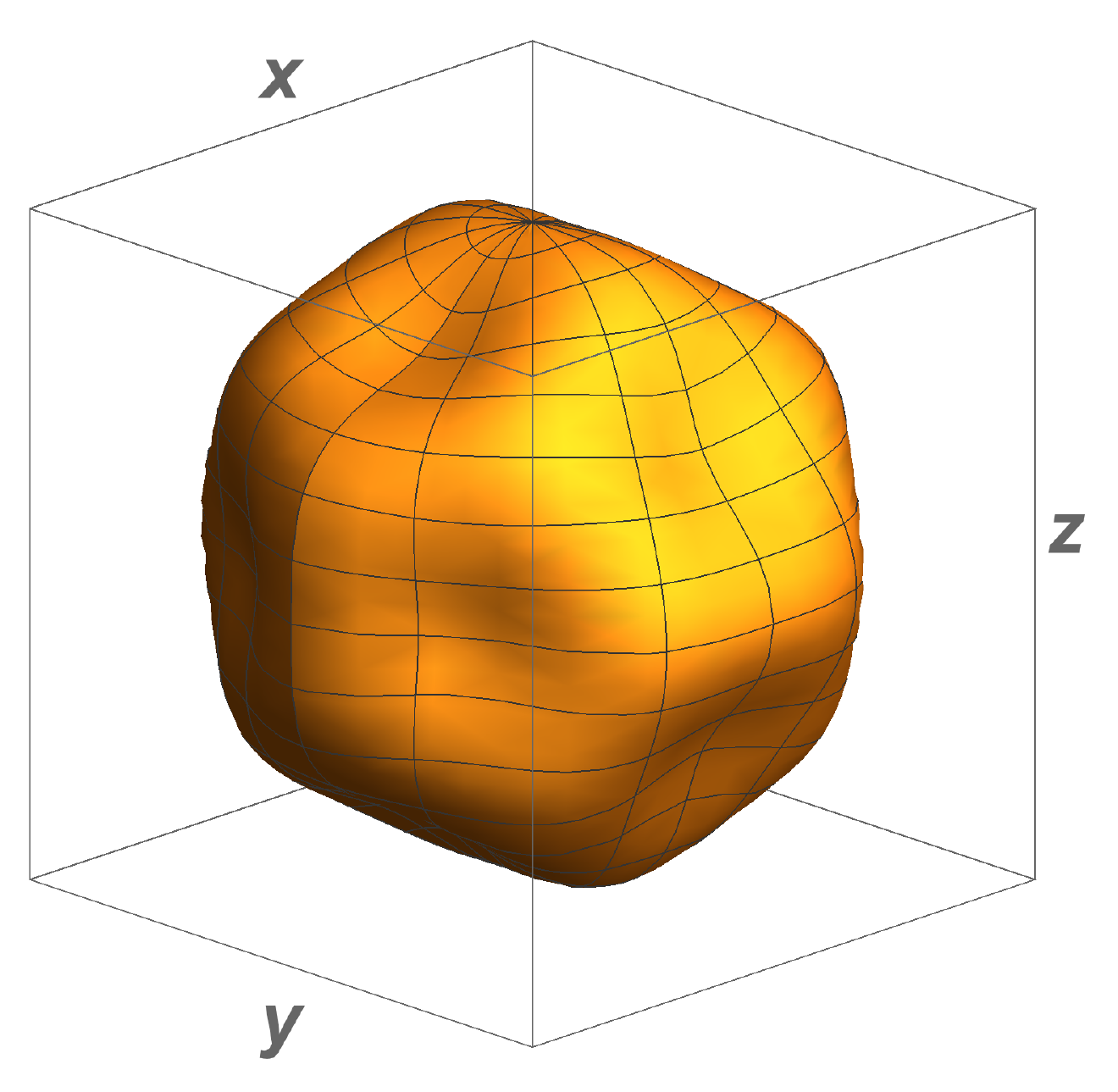}  
} \hspace{1cm}
\subfloat{\includegraphics[width=5.5cm]{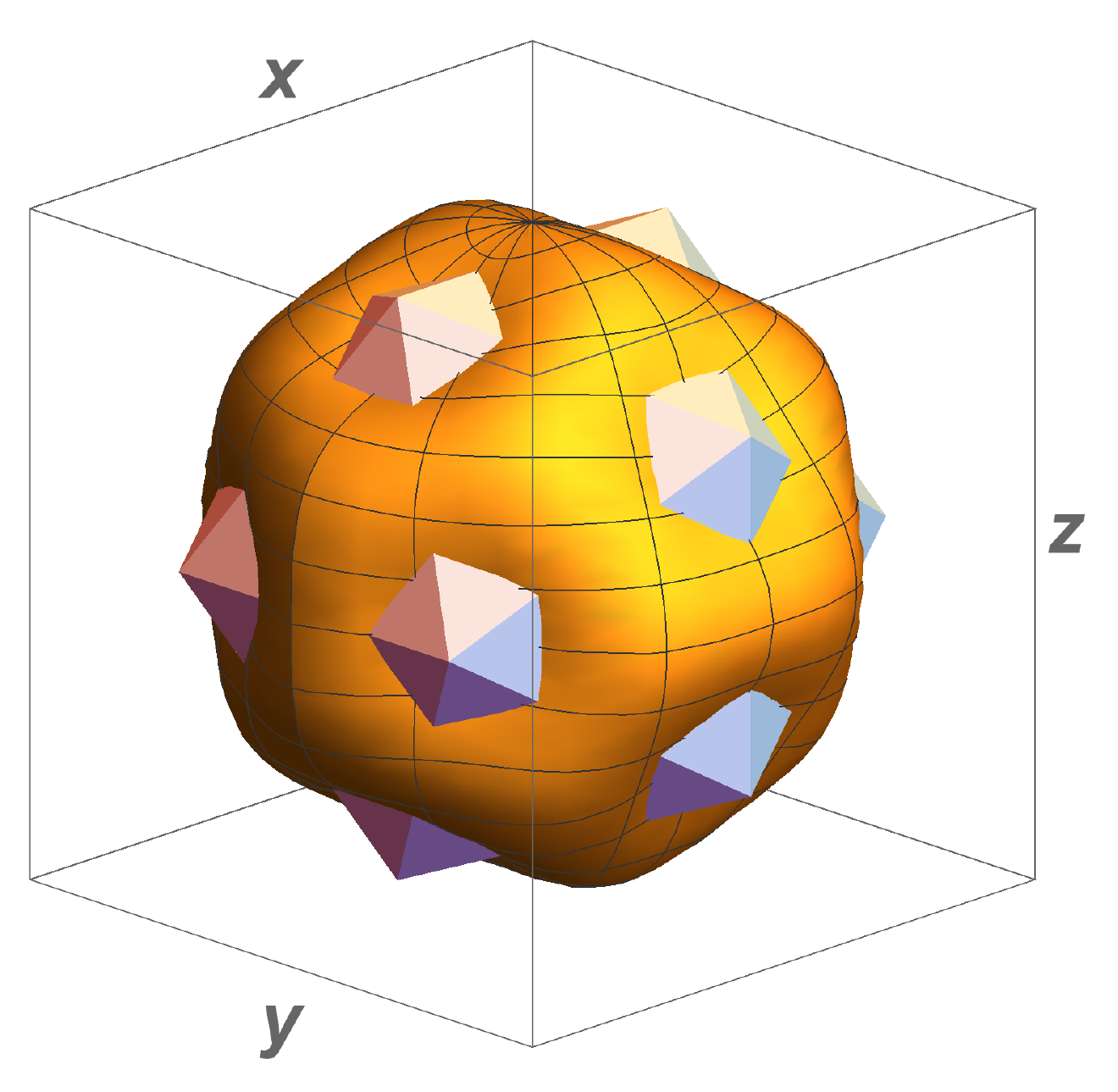} 
}
\caption{\label{full ++} \small $ \langle \gamma^+ \gamma^+ \rangle = \langle \gamma^- \gamma^- \rangle $ for $\Delta c^2_\gamma = 20\%$.}
\end{figure}

We plot the entries of $M$ in figs.~\ref{++} and \ref{+-}; for $M^{+-}$ and $M^{-+}$, we only plot their absolute values. In these figures we also show the orientation of the icosahedron underlying our model.
We can clearly see that the spikes of $M^{++} = M^{--}$ are aligned with the vertices of the underlying icosahedron, and the spikes of $|M^{+-}| = |M^{-+} |$ are aligned with its edges. Clearly, the geometry  of our icosahedron is imprinted  in the tensor modes' two-point functions.  For illustrative purposes, in fig.~\ref{full ++} we plot the directional dependence of the full $\langle \gamma^+ \gamma^+ \rangle$ two-point function for $\Delta c^2_\gamma = 20\%$.

Finally, we can compute directly the two-point function of the tensor field $\gamma_{ij}(\vec k)$, without decomposing this into separate helicities; such a two-point function is insensitive to the ambiguities (and singularities) of the polarization tensors. Using the identities of sect.~\ref{identities}, we find:
\be \label{tensor tensor}
\langle \gamma_{ij}^1 \gamma_{kl}^2 \rangle = \big[2 P^{TT}_{(ij),(kl)} - 3 \Delta c^2_\gamma \, P^{TT}_{(ij),(mn)} P^{TT}_{(kl),(op)} \, T_6^{mnopqr} \, \hat k_q  \hat k_r \big] \cdot \langle \gamma_1 \gamma_2 \rangle_0 \; ,
\ee
where $P^{TT}$ is the projector onto symmetric, transverse, traceless tensors,
\be
P^{TT}_{(ij),(mn)} = \sfrac12 \big( P_{im} P_{jn} + P_{in} P_{jm} - P_{ij} P_{mn} \big) \; .
\ee

\section{Concluding remarks and outlook}
Our results show that, in the presence of higher-derivative interactions, the tensor modes of icosahedral inflation generically acquire an anisotropic power spectrum, with features aligned with the underlying icosahedral structure. Moreover, they also acquire a mixed correlator between the two different helicities, which is forced to vanish in all isotropic models because of rotational invariance. These effects are perturbatively small within the regime of validity of the effective theory. How small? If we take as an illustrative example the higher derivative interaction suggested in \cite{Kang:2015aa},
\be \label{hd}
{\cal L}_{\rm h.d.} \sim \frac{1}{M^2} (R^{\mu\nu\rho\sigma} \partial_\mu \phi^I \partial_\nu \phi^J \partial_\rho \phi^K \partial_\sigma \phi^L )^3 \; ,
\ee
where $M$ is some mass scale and suitable contractions with our $T_6^{IJKLMN}$ are understood, we discover that at the quadratic level in $\gamma_{ij}$ we have two effects \cite{us2}:
\begin{enumerate}
\item An anisotropic correction to the tensors' propagation speed  of order
\be
\Delta c_\gamma^2 \sim \frac{H^4}{M^2 \mpl^2} \; ;
\ee
\item 
A higher-derivative correction to the kinetic Lagrangian schematically of the form
\be
\frac{H^2}{M^2} (\partial^2 \gamma)^2 \; .
\ee
\end{enumerate}
The effective field theory breaks down when the latter starts competing with the lowest order kinetic energy we get from Einstein-Hilbert, $\mpl^2 (\partial \gamma)^2$.
For the effective theory to be valid at least up to frequencies of order Hubble, we thus need
\be
\frac{H^4}{M^2 \mpl^2} \lesssim 1  \quad \Rightarrow \quad \Delta c_\gamma^2 \lesssim 1 \; .
\ee

Since $\Delta c^2_\gamma$ measures directly the relative importance of our anisotropic contributions to the tensors' two-point functions, we see that such contributions are expected to be somewhat smaller than one (in relative terms), but not necessarily much smaller.
Of course, if $M^2$ is so high that the effective theory is still valid at frequencies much higher than Hubble, our effects will be much smaller than one.

Notice that, for the same reason as mixed $+-$ tensor correlators are allowed in our model, mixed scalar-tensor correlators are also  allowed. Given the smallness of the tensor-to-scalar ratio, these could be bigger that the tensor spectra themselves, and thus offer an interesting observational opportunity. We leave computing them in detail for future work, but here we can sketch an order of magnitude estimate.

Let's take again eq.~\eqref{hd} as an illustrative example. Expanding to bilinear order in $\gamma_{ij}$ and in the phonon field $\pi^i$ we can get a three-derivative term of the form
\be
\frac{H^4}{M^2} T_6^{ijklmn} \partial_i  \pi_j \, \partial_k \partial_l \gamma_{mn} \sim \Delta c_\gamma^2 \mpl^2 \, T_6^{ijklmn} \partial_i  \pi_j \, \partial_k \partial_l \gamma_{mn} \; . 
\ee
Combining this with the lowest order $\gamma$ and $\pi$ kinetic terms \cite{Endlich:2012pz}---evaluated for simplicity at frequencies of order Hubble and for a relativistic phonon speed, $c_L\sim1$---we get schematically
\be 
\mpl^2 \big[ (\partial \gamma)^2 + \epsilon H^2 (\partial \pi)^2 + \Delta c_\gamma^2 T_6 \, \partial \pi \partial^2 \gamma  \big] 
\ee
Renormalizing the phonon field as $\tilde \pi \equiv \sqrt \epsilon H \cdot \pi$, this becomes
\be
\mpl^2 \big[ (\partial \gamma)^2 + (\partial \tilde \pi)^2 + \frac{\Delta c_\gamma^2}{\sqrt \epsilon H} \,  T_6 \, \partial \tilde \pi \partial^2 \gamma  \big] 
\ee
We thus expect an anisotropic mixed $\tilde \pi$-$\gamma$ correlator of order
\be
\langle \tilde \pi \gamma \rangle \sim \frac{\Delta c_\gamma^2}{\sqrt \epsilon} \,  \langle \tilde \pi  \tilde \pi \rangle \sim  \frac{\Delta c_\gamma^2}{\sqrt \epsilon} \,   \langle \gamma \gamma \rangle
\ee
Recalling that $\zeta$ is related to $\pi$ by $\zeta = \frac13 \vec \nabla \cdot \vec \pi$ \cite{Endlich:2012pz}, we finally get
\be
\langle \zeta \gamma \rangle \sim \Delta c_\gamma^2 \langle \zeta \zeta \rangle \sim \frac{\Delta c_\gamma^2}{\epsilon} \langle \gamma \gamma \rangle
\ee
In principle, this could be of the same order of or even larger than the tensor spectrum, as long as $\Delta c_\gamma^2 \gtrsim \epsilon$.

\section*{Acknowledgements}

We would like to thank Lam Hui and Austin Joyce for useful discussions. 
Our work is partially supported by the DOE under contract no.~DE-FG02-11ER41743 and by the Kwanjeong Educational Foundation.

\section*{Appendix}

\appendix

\section{Beyond perturbation theory}
In this Appendix, we compute the tensor two-point functions without relying on a perturbative analysis. Let us rewrite $\lagr_\text{int}$ in matrix form:
\be
\begin{split}
S_\text{int} 
= -\fr{\mpl^2 \Delta c_\gamma^2}{8} \int \fr{d\tau d^3 k}{(2\pi)^3} \, a^2 k^2 \gamma^\dagger(\vec{k}) M(\vec{k}) \gamma(\vec{k}) \, ,
\end{split}
\ee
where
\be
\gamma(\vec{k}) \equiv 
\begin{pmatrix}
\gamma^+(\vec{k})
\\
\gamma^-(\vec{k})
\end{pmatrix}
\ee
and $M(\vec k)$ is the  matrix that we defined in \eqref{M}.
$M(\vec{k})$ is hermitian and thus diagonalizable. However, the unitary transformation needed to diagonalize it depends on the direction of $\vec k$. We will thus use the diagonal form of $M$ as an intermediate step to compute the tensor modes' two-point functions, but we will eventually express these in the original $\gamma^{\pm}$ basis. 
Defining
\be
\gamma_d(\vec{k}) \equiv
\begin{pmatrix}
\gamma^1(\vec{k})
\\
\gamma^2(\vec{k})
\end{pmatrix}
= \fr{1}{\sqrt{2}}
\begin{pmatrix}
\gamma^-(\vec{k}) - M^{+-}(\vec{k})\gamma^+(\vec{k}) / |M^{+-}(\vec{k})|
\\
\gamma^-(\vec{k}) + M^{+-}(\vec{k})\gamma^+(\vec{k}) / |M^{+-}(\vec{k})|
\end{pmatrix} \, ,
\ee
we have
\be
\gamma^\dagger(\vec{k}) M(\vec{k}) \gamma(\vec{k}) = \gamma_d^\dagger(\vec{k})
\begin{pmatrix}
M^{++}(\vec{k}) - |M^{+-}(\vec{k})| & 0
\\
0 & M^{++}(\vec{k}) + |M^{+-}(\vec{k})| 
\end{pmatrix}
\gamma_d(\vec{k}) \, .
\ee

One can now compute the two-point functions of $\gamma^1(\vec{k})$ and $\gamma^2(\vec{k})$ in the usual way, since these are decoupled. One can then go back to the $\gamma^\pm$ basis and obtain the exact two-point functions
\begin{multline}
\l \langle \gamma^+_1 \gamma^+_2 \r \rangle  = \l \langle \gamma^-_1 \gamma^-_2 \r \rangle
\\
= \fr{H^2}{\mpl^2}\fr{1}{2k_1^3}\l( \fr{1}{\l(1 + \fr{\Delta c_\gamma^2}{2}(M^{++}(\vec{k}_1) - |M^{+-}(\vec{k}_1)|)\r)^\fr{3}{2}} +  \fr{1}{\l(1 +\fr{\Delta c_\gamma^2}{2}(M^{++}(\vec{k}_1) + |M^{+-}(\vec{k}_1)|)\r)^\fr{3}{2}} \r) \, ,
\end{multline}
and
\begin{multline}
\l \langle \gamma^-_1 \gamma^+_2 \r \rangle =  \l \langle \gamma^+_1 \gamma^-_2 \r \rangle ^* = \fr{H^2}{\mpl^2}\fr{M^{+-}(\vec{k}_1)}{2k_1^3 |M^{+-}(\vec{k}_1)|}
\\
\times \l( - \fr{1}{\l(1 + \fr{\Delta c_\gamma^2}{2}(M^{++}(\vec{k}_1) - |M^{+-}(\vec{k}_1)|)\r)^\fr{3}{2}} + \fr{1}{\l(1 + \fr{\Delta c_\gamma^2}{2}(M^{++}(\vec{k}_1) + |M^{+-}(\vec{k}_1)|)\r)^\fr{3}{2}} \r ) \, ,
\end{multline} 
where we omitted the standard $(2\pi)^3 \delta^3 (\vec{k}_1 + \vec{k}_2)$ factors. 
It is immediate to check that, once expanded to first order in $\Delta c_\gamma^2$, the above expressions reduce precisely to our perturbative results, eqs.~\eq{eq:2ptpp_done} and \eq{eq:2ptmp_done}.




\section{Ambiguities and singularities of polarization tensors}\label{phases}

As usual, it is convenient to define our helicity-$\pm2$ polarization tensors as tensor products of helicity-$\pm1$ polarization vectors,
\be
\epsilon_{ij}^{2 h}(\vec k) \equiv \sqrt{2} \, \epsilon_{i}^{h}(\vec k) \, \epsilon_{j}^{h}(\vec k) \; , \qquad h=\pm 1
\ee
(the prefactor is chosen for consistency with the standard normalization for $\vec \epsilon \, ^{h}(\vec k)$ and with \eqref{epsilon}).
Then, without loss of generality, we can just study the properties of  polarization vectors $\vec \epsilon \, ^{h}(\vec k)$ of definite helicity.

For given $\vec k$ and $h$, the defining properties of $\vec \epsilon \, ^{h}(\vec k)$ are transversality, normalization, and helicity:
\be
\vec k \cdot \vec \epsilon \, ^{h}(\vec k) = 0 \; , \qquad \vec \epsilon \, ^{h}(\vec k) \cdot \vec \epsilon \, ^{h \, *}(\vec k) = 1 \; , \qquad R_{\hat k}(\alpha) \cdot \vec \epsilon \, ^{h}(\vec k) = e^{i h \alpha} \, \vec \epsilon \, ^{h}(\vec k) \; ,
\ee
where $R_{\hat k}(\alpha)$ is a counterclockwise rotation of angle $\alpha$ around $\hat k$. These conditions  determine $\vec \epsilon \, ^{h}(\vec k)$ only up to its phase, as clear from the fact that they are all invariant under a rephasing of $\vec \epsilon \, ^{h}(\vec k)$.

Now, this arbitrary phase can be be varied at will as a function of $\vec k$. To simplify one's life, one can try and enforce certain relationships among the $\vec \epsilon \,$'s associated with different $\vec k$'s. For instance, it is technically useful to impose
\be
\vec \epsilon \, ^{h}(-\vec k) = \vec \epsilon \, ^{h \, *}(\vec k) = \vec \epsilon \, ^{-h}(\vec k) \; , \qquad \vec \epsilon \, ^{h}(\vec k) = \vec \epsilon \, ^{h}(\hat k) \; ,
\ee
where the latter means that all $\vec k$'s pointing in the same direction share the same polarization vectors. But there is no preferred relationships among the polarization vectors for $\vec k$'s pointing in different directions: the $\hat k$-dependence of the phase of $\vec \epsilon \, ^{h}(\hat k)$ is completely arbitrary. 

Perhaps surprisingly, what is not arbitrary is that such a phase, as a function of $\hat k$, must be {\em singular} somewhere. To see this, consider changing basis and going from circular ($+$ and $-$) to linear polarizations (1 and 2), which are defined in terms of two orthonormal, transverse, real vectors $\vec \epsilon \, ^{1,2}(\hat k)$. The fact that these depend just on $\hat k $ and that they are transverse to it, means that they are tangent vector fields on the unit  two-sphere. But there is no way to have a regular vector field on a two-sphere that does not vanish anywhere. Since $\vec \epsilon \, ^{1,2}(\hat k)$ cannot vanish anywhere---they are normalized to one---, they cannot be regular functions of $\hat k$.

For example, consider the standard procedure, whereby one defines $\vec \epsilon \, ^{h}(\hat k)$ for $\hat k $ parallel to the $z$-axis as
\be
\vec \epsilon \, ^{\pm1}(\hat z) = \frac{1}{\sqrt{2}} \big(1,\pm i, 0 \big) \; ,
\ee
and then, for any other $\hat k$, one picks a standardized rotation $R_{\hat z \to \hat k}$ that rotates $\hat z$ to $\hat k$, and defines $\vec \epsilon \, ^{h}(\hat k)$ as
\be
\vec \epsilon \, ^{h}(\hat k) \equiv R_{\hat z \to \hat k} \cdot \vec \epsilon \, ^{h}(\hat z) \; .
\ee
There is no preferred choice for $R_{\hat z \to \hat k}$. Different choices define the same $\vec \epsilon \, ^{h}(\hat k)$ up to a $\hat k$-dependent phase, as expected from the discussion above. Let's choose for instance
\be
R_{\hat z \to \hat k} = R_z(\phi) \cdot R_y(\theta) \; ,
\ee
where $R_i(\alpha)$ implements a counterclockwise rotation of angle $\alpha$ around the $i$ axis, and $(\theta, \phi)$ are the polar and azimuthal angles of $\hat k$. Everything looks regular in the definition of $\vec \epsilon \, ^{h}(\hat k) $ above, but one must recall that $\phi$ is not well-defined at the poles. And so, for instance, if one approaches the North pole along a meridian of longitude $\phi$, one gets
\be
\vec \epsilon \, ^{h}(\hat k \simeq \hat z) \simeq e^{ i h \phi} \, \frac{1}{\sqrt{2}} \big(1,\pm i, 0 \big) \; , 
\ee
which is clearly discontinuous at the pole. Of course, the same happens at the South pole.

Mutatis mutandis, all of the above applies to our polarization tensors.


\section{Useful identities for tensor products of polarization tensors}\label{identities}

The arbitrary and singular phases discussed in Appendix \ref{phases} cancel if we consider the tensor product of a polarization vector or tensor with its complex conjugate. As a consequence, such tensor products are completely unambiguous, and in fact can be rewritten in a manifestly covariant form, which simplifies considerably the derivations of eqs.~\eqref{M++}, \eqref{M+-}, and \eqref{tensor tensor}. 

To begin with, consider the helicity $h=\pm1$ polarization vector for $\hat k$ aligned with $\hat z$,
\be
\vec \epsilon \, ^{\pm1}(\hat z) \propto \frac{1}{\sqrt{2}} \big(1,\pm i, 0 \big) \; ,  \label{z}
\ee
where the proportionality factor is just a phase. We have
\be
\epsilon_i^{h}(\hat z) \epsilon_j^{h \, *}(\hat z) =  \frac{1}{{2}} \left( \begin{array}{ccc}
1 & - i h & 0 \\
+i  h & 1 & 0 \\
0& 0 & 0 
\end{array}\right) = \frac12\big(\delta_{ij} - \hat z_i \hat z_j - i h \epsilon_{ijk} \, \hat z_k \big) \; . \label{e e*}
\ee

For any other $\hat k$, the polarization vectors are just rotated versions of \eqref{z}. Obviously, rotating  \eqref{e e*}
we get
\be
\epsilon_i^{h}(\hat k) \epsilon_j^{h \, *}(\hat k) = \sfrac12 \big( P_{ij}(\hat k )- i h \varepsilon_{ijk} \, \hat k_k \big) \; ,
\ee
where $P_{ij}(\hat k)$ is the transverse projector
\be
P_{ij}(\hat k) = \delta_{ij} - \hat k_i \hat k_j \; .
\ee

We can now use this result for our polarization tensors, which are just tensor products of polarization vectors. We get
\begin{align}
\epsilon_{ij}^{2h}(\hat k) \epsilon_{kl}^{2h \, *}(\hat k) & = \sfrac12(P_{ik}- i h \epsilon_{ikm} \, \hat k_m)(P_{jl}- i h \epsilon_{jln} \, \hat k_n) \nonumber \\
& = \sfrac12\big[ P_{ik} P_{jl} + P_{il} P_{jk} - P_{ij} P_{kl}  - i h \big( P_{ik} \epsilon_{jln} + P_{jl} \varepsilon_{ikn}\big) \hat k_n \big] \; ,  \label{tensor product}
\end{align}
where for notational simplicity we dropped the argument of $P$. This expression is all we need to compute $M^{++} = M^{--}$ in \eqref{M++}.

For $|M^{+-}|^2$ (eq.~\eqref{M+-}) and for $\langle \gamma_{ij} \gamma_{kl} \rangle$ (eq.~\eqref{tensor tensor}), we need, schematically, $\epsilon^{2h} \epsilon^{2h \, *} \epsilon^{2h'} \epsilon^{2h' \, *}$, where all the $\epsilon$'s are at the same $\hat k$. 
More specifically, for the former we need the sum
\be \label{single sum}
\sum_{h=\pm1} \epsilon_{ij}^{2h} \epsilon_{kl}^{2h \, *} \epsilon^{2h}_{mn} \epsilon^{2h \, *}_{op} = \sum_{h=\pm1} \epsilon^{4h}_{ijmn} \epsilon^{4h \, *}_{klop} \; ,
\ee
where the $\epsilon^{4h}$'s are helicity-$\pm 4$ polarization tensors, while for the latter we need the double sum
\be \label{double sum}
\sum_{h, h'=\pm1} \epsilon_{ij}^{2h} \epsilon_{kl}^{2h \, *} \epsilon^{2h'}_{mn} \epsilon^{2h' \, *}_{op} = \bigg( \sum_{h =\pm1} \epsilon_{ij}^{2h} \epsilon_{kl}^{2h \, *} \bigg) \bigg( \sum_{h' =\pm1} \epsilon_{mn}^{2h'} \epsilon_{op}^{2h' \, *} \bigg) \; . 
\ee

All of these sums can be rewritten in terms of projectors. As clear from \eqref{tensor product}, each sum on the r.h.s~of \eqref{double sum} is just (twice) 
the standard symmetric-transverse-traceless projector for two-index tensors:
\be
 \sum_{h =\pm1} \epsilon_{ij}^{2h} \epsilon_{kl}^{2h \, *} = 2 P^{TT}_{(ij),(kl)} \equiv \big( P_{ik} P_{jl} + P_{il} P_{jk} - P_{ij} P_{kl}  \big) \; .
\ee
This is to be expected: for any given $\vec k$, in the vector space of complex two-index tensors with scalar product $A_{ij}^*B_{ij}$, our polarization tensors $\epsilon_{ij}^{2h}$ furnish an orthonormal basis for the subspace of symmetric, transverse, traceless tensors. The extra factor of $2$ above stems from the normalization condition in \eqref{epsilon}.

Then, we can immediately generalize this result to \eqref{single sum}: in the vector space of complex four-index tensors with scalar product $A_{ijkl}^*B_{ijkl}$, the two  polarization tensors $\epsilon_{ijkl}^{4h}$ furnish an orthonormal basis for the subspace of totally symmetric, transverse, traceless tensors. Thus, the sum in \eqref{single sum} is simply the projector onto this subspace, with an extra overall factor of $4$, again because of normalization:
\be
\sum_{h=\pm1} \epsilon_{ij}^{2h} \epsilon_{kl}^{2h \, *} \epsilon^{2h}_{mn} \epsilon^{2h \, *}_{op} = 4 P^{TT}_{(ijmn), (klop)}   \; .
\ee

When such a projector is contracted with two totally symmetric tensors with indices $(ijmn)$ and $(klop)$---the case relevant for eq.~\eqref{M+-}---, we can forget about  symmetrizing it  and replace it with a much simpler operator, in which we project all the indices and subtract just the traces with the correct combinatoric factors. In particular, the traceless part of a four-index tensor $A_{ijmn}$ that is already totally symmetric and transverse is 
\be
\bar A_{ijmn} = A_{ijmn}- \sfrac16 \big(P_{ij} A_{kkmn} + \mbox{5 perms.}\big) + \sfrac1{24} \big(P_{ij} P_{mn} + \mbox{2 perms.}\big) A_{kkll} 
\ee
And so, our projector can be replaced by
\be
P^{TT}_{(ijmn), (klop)} \to P_{ik} P_{jl} P_{mo} P_{np} -  P_{ij} P_{kl} P_{mo} P_{np} + \sfrac18 P_{ij} P_{kl} P_{mn} P_{op} \; .
\ee

\bibliographystyle{utphys_jonghee}
\bibliography{icosahedral_tensor}{}

\end{document}